\documentclass{ametsocV6.1}

\usepackage{etoolbox}

\nolinenumbers  

\usepackage{lineno}

\makeatletter
\AtBeginDocument{%
  \nolinenumbers
  \let\internallinenumbers\relax 
}
\makeatother

\usepackage{amsmath,amsfonts,amssymb,bm}
\usepackage{mathptmx}
\usepackage{newtxtext}
\usepackage{newtxmath}

\title{Winter Precipitation Type Diagnosis and Uncertainty Quantification with a Physically Consistent Machine Learning Method}

\authors{
Charlie Becker\aff{1}\correspondingauthor{Charlie Becker, cbecker@ucar.edu},
David John Gagne II\aff{1},
Julie Demuth\aff{1},
John S. Schreck\aff{1},
Jacob Radford\aff{3},
Gabrielle Gantos\aff{1},
Eliot Kim\aff{1},
Dhamma Kimpara\aff{2},
Sophia Reiner\aff{4},
Justin Willson\aff{1},
Christopher D. Wirz\aff{1,5}}

\affiliation{
\aff{1}{NSF National Center for Atmospheric Research, Boulder, CO, USA}\\
\aff{2}{Department of Computer Science, University of Colorado, Boulder, CO, USA}\\
\aff{3}{Cooperative Institute for Research in the Atmosphere, Colorado State University, Fort Collins, CO, USA}\\
\aff{4}{Department of Computer Science, University of Wisconsin-Madison, Madison, WI, USA}\\
\aff{5}{Agricultural Leadership, Education, and Communication, University of Illinois Urbana-Champaign, Urbana, IL, USA}\\
}

\abstract{
Accurately forecasting winter precipitation type and its transitions is critical for high-impact decision making. However, existing methods struggle in thermodynamically ambiguous regimes, and most do not quantify forecast uncertainty from a single model run. We developed an evidential neural network that predicts calibrated probabilities for four winter precipitation types (rain, snow, freezing rain, and ice pellets) along with epistemic uncertainty estimates at the computational cost of a standard neural network. The model was trained on quality-controlled and curated observations from the crowd-sourced mPING dataset paired with vertical thermodynamic profiles from the NOAA Rapid Refresh model analyses. Rigorous physical quality control removed thermodynamically implausible reports. Bulk evaluation against held-out mPING observations from June 2020 through June 2022 shows the ML model outperforms area-based deterministic methods in success ratio for freezing rain and ice pellets while maintaining comparable or better performance for rain and snow. A reduced freezing rain probability of detection reflects genuinely ambiguous thermodynamic environments rather than a uniform model deficiency and is more robustly represented through the full probability distribution than through the dominant predicted class alone. Thermodynamic regime analysis demonstrates that model prediction errors are physically structured and concentrated in interpretable regions of diagnostic space consistent with the known difficulty of freezing rain and ice pellet discrimination. We further demonstrate the model's physical consistency and operational utility through two contrasting mid-western U.S. winter storm case studies and an interactive visualization tool that enables dynamic interrogation of model predictions and uncertainty in real time.}

\begin{document}
\maketitle

\section{Introduction}

Accurate forecasts of winter precipitation type are essential for effective decision-making during high-impact winter storms, yet the societal consequences of different precipitation types and errors in their prediction vary considerably \citep{Lazo2020-rv}. Freezing rain and ice pellets present particular hazards: even small accumulations of freezing rain can make roads and infrastructure dangerously slick, while ice pellets can reduce visibility and create hazardous travel conditions that are difficult to anticipate \citep{Black2015-rd}. The transition between precipitation types adds further complexity, as missing the onset of frozen precipitation can leave roads untreated and lead to accidents and stranded vehicles \citep{Black2015-rd}. Beyond road safety, precipitation type affects aviation operations, power infrastructure, agriculture, and public event planning, with decision makers across these sectors relying on forecasts that convey not just the expected type but the uncertainty and timing of transitions. Addressing these societal impacts requires not only accurate precipitation type forecasts but also reliable quantification of forecast uncertainty, particularly in the thermodynamically ambiguous regimes where transitions between precipitation types are most likely and the consequences of forecast errors are greatest.

Forecasters currently rely on a variety of post-processing methods to determine precipitation type (hereafter referred to as p-type) that can be applied directly to numerical weather prediction (NWP) output or observed atmospheric profiles. Significant contributions were developed in the 1990s \citep{Baldwin:1993, Bourgouin:2000, Ramer:1993, Cantin:1993}, many of which have been modified \citep{Benjamin2016-xz, Birk2021-ch, Manikin_2005} and continue to be used today. The area-based methods \citep{Bourgouin:2000, Baldwin:1993} use the area of a thermodynamic vertical profile (either dry-bulb temperature or wet-bulb temperature) above or below the freezing point in a decision tree as the basis for p-type. Other implicit decision tree-based methods include the ice-fraction method \citep{Ramer:1993}, which accounts for the precipitation ice-fraction at the precipitation generation layer for a decision tree or the partial thickness method \citep{Cantin:1993}, where decisions are made using the thermodynamic profile within different geopotential heights. P-type can also be determined explicitly through the use of more modern microphysical schemes \citep{Benjamin2016-xz, Reeves2016-vf, Thompson:2008} which can represent clouds and supercooled water droplets. Despite the improvements over the last three decades, no single method has unambiguously been declared superior, and there is evidence that each algorithm may perform well in some situations and poorly in others \citep{Reeves2023-hp}. This lack of consensus on a dominant algorithm has led to some NWP ensembles, such as the Short-Range Ensemble Forecast (SREF), to use a mix of postprocessing p-type methods for individual ensemble members to determine the most likely p-type \citep{Manikin_2005}. 

Forecasting the timing, transition, and duration of these winter events is challenging, particularly when the near-surface temperature is close to 0 $^{\circ}$C \citep{minder-ptype}. Analyses of near-surface temperatures can be erroneous by upwards of 4 $^{\circ}$C \citep{coniglio:2007} due to data assimilation uncertainties, yet p-type can be altered by low-level changes of less than 0.5$^{\circ}$C \citep{Theriault:2010}. Additionally, biases in NWP models can arise from simplifying assumptions in microphysics parameterizations, insufficient vertical grid spacing to resolve shallow cold (warm) layers, or inconsistencies with a land surface model and the latent heat exchange \citep[e.g.,][]{lackman-ptype}. These biases, in addition to uncertainty throughout the thermodynamic profile, make consistent, accurate discrimination between frozen precipitation types inherently uncertain \citep{Reeves2014-nv, Ralph:2005, Stewart:2015, lackman-ptype}. The discrimination between ice pellets (sleet) and freezing rain is particularly challenging \citep{Bourgouin:2000, Manikin_2005, Reeves2014-nv, Elmore2011-jo} due to the similarities in their respective temperature profiles. Another challenge is how to best represent mixed precipitation events when multiple precipitation types are occurring simultaneously in a given area. 

Real-world thermodynamic complexities that are simplified or poorly represented in physics-based models and diagnostic methods have motivated the use of machine learning to learn these complexities directly from observations. \cite{McGovern2017-av} used the Meteorological Phenomena Identification Near the Ground (mPING, \cite{MPING}) data as observations to train four separate random forests, each corresponding to a specific p-type. \cite{Pham2023} used in-situ daily observations to classify between rain and snow for better downstream hydrologic estimates. \cite{Zhuang2024} used quality controlled METAR observations in conjunction with ERA5 reanalysis data to identify biases such as biases in elevation when using pressure level data and a severe freezing drizzle bias. \cite{Filipiak_2023} used the Community Collaborative Rain, Hail and Snow Network (CoCoRHaS, \cite{cocorhas}) dataset in conjunction with the New York State Mesonet instrumentation \citep{nym} data to predict p-type. As evidenced by these examples all using a different data source for training targets, there is an inherent challenge in using ML as a post-processing method to accurately predict p-type: which data source(s) should be chosen for training targets and verification?

The disparity in observational targets highlights the potential strengths and shortcomings of each of these datasets, which pertain to factors such as spatial and temporal resolution and availability, accuracy, and types of precipitation observations recorded. Crowd-sourced data sets, such as mPING and CoCoRHaS, often have denser spatial coverage than in-situ instrumentation but lack sampling consistency and can suffer from high variance in the assessment skills of the observers. In-situ networks such as the Automated Surface/Weather Observing Systems (ASOS / AWOS) that are deployed at most airports across the United States generally have high frequency sampling but sparse spatial coverage and lack the ability to observe ice pellets with an automated sensor, resulting in only sites with human observers reporting this p-type. Additionally, there are biases and uncertainties in the observations themselves \citep{Reeves2016-vf, Landolt2019-yu} which is an ongoing challenge for both ML methods and verification. In this study, we utilize mPING as our primary observational dataset, motivated by its dense spatial coverage, availability of ice pellet reports unavailable from automated ASOS sensors, and its large archive spanning multiple winter seasons. While crowd-sourced observations introduce the reliability challenges described above, we address these through rigorous physical quality control (Section 3) that substantially reduces thermodynamically implausible reports and improves the physical consistency of model training and verification.

The aforementioned post-processing methods to determine p-type can be broadly split into two types: heuristic and statistical methods. The heuristic, tree-based methods have advantages by generally being easy to implement, easy to interpret, and quickly tuned to a set of observations. The statistical / ML methods are often more difficult to implement and interpret, but they offer the advantages of higher potential accuracy and uncertainty quantification through probabilistic output. Accurately conveying these uncertainties in winter p-type forecasting can support forecasters' decision-making and facilitate more nuanced communication with their core partners (e.g., emergency managers, transportation officials) leading up to and during challenging events \citep{Novak2023-th, rogers2023nws, Joslyn2012-kv}. Researchers and model developers also benefit by potentially identifying limits of predictability or areas where more data collection would drive improvement. For example, irreducible uncertainty as a result of the training data itself (aleatoric uncertainty) can highlight data overlaps that are unlikely to be reduced with more data collection, whereas reducible uncertainty within the model solution space (epistemic uncertainty) may highlight areas where more targeted data collection could improve model performance. Furthermore, providing some interpretation of the model predictions and uncertainties through regime-based physical analysis and interactive tools may increase the trustworthiness of the model and provide a better framework for feedback.

There are numerous techniques to help quantify uncertainty including ensemble approaches such as Monte Carlo dropout, deep ensembling, and quantile regression (see \cite{Haynes:2023} for an overview of these methods in an Earth Systems context). However, all of these methods require extra computation to calculate epistemic uncertainty by either sampling or ensemble strategies. A more recent approach, evidential deep learning \citep{Sensoy2018-jy, amini-evi}, predicts second order distributions which can explicitly estimate epistemic uncertainties with a single model. The evidential models are as computationally efficient as a traditional neural network with the only architectural change required being a custom loss function. A detailed review of evidential models in both the classification and regression setting for Earth system science can be found in \cite{evidential-schreck}. Additionally, we have developed MILES-GUESS, an open source software package for uncertainty quantification (UQ), including evidential models, that was the software platform used for this research \citep{miles-guess}. 

The focus of this paper is to combine rigorous quality control with an evidential neural network to better capture the uncertainties and explainability associated with the machine learning of winter precipitation type. Specifically, we aim to address two primary questions: (1) how can quality control (QC) and data curation of crowd-sourced data affect model performance and physical explainability and (2) what physical regimes are associated with various levels of p-type uncertainty? We demonstrate our methods through bulk statistical evaluation against quality controlled mPING observations, the Rapid Refresh \citep[RAP;][]{Benjamin2016-mj} NWP model, and both the original and modified Bourgouin area methods, complemented by thermodynamic regime analysis of model prediction errors, two mid-western U.S. winter storm case studies, and an interactive visualization tool that enables dynamic interrogation of model predictions and uncertainty.

\section{Data}

\subsection{Meteorological Input Data}

Most decision-tree based precipitation type methods that are currently used depend on some form of the vertical thermodynamic profile in the atmosphere, although the variables and levels at which decisions are made differ by method. Our approach is to use the vertical thermodynamic profile in a neural network and attempt to implicitly learn the important non-linear relationships in the profile as it relates to precipitation type. By accounting for the entire profile (to the height above ground we specify), we can aim to capture the different complex conditions that relate to precipitation type that cannot be accounted for by simple decision thresholds.  

We derive our input variables from the vertical profiles taken from the National Centers for Environmental Information (NCEI) archive of RAP analyses, which stores profiles of temperature, relative humidity, and winds on 37 pressure levels every 25 hPa from 100 to 1000 hPa at a 13 km grid spacing \citep{Benjamin2016-mj}. We convert the relative humidity to dew point and select the dew point, temperature, and U and V wind components throughout the profile. Dry and saturated layers can share the same wet-bulb temperature yet represent thermodynamically distinct environments with different implications for hydrometeor phase, so providing temperature and dewpoint separately rather than a derived wet-bulb temperature preserves the model's ability to distinguish between these cases. We linearly interpolate all profile variables from pressure levels to height above ground level at 250 meter intervals for several reasons: meteorological features relevant to precipitation type such as the warm nose and freeze layer, appear at more consistent vertical positions in height coordinates, simplifying the relationships the model must learn; pressure-level fields provide effectively coarser vertical resolution over higher terrain where fewer levels fall within the lowest few kilometers of the atmosphere; and a fixed height grid ensures consistent input structure regardless of surface elevation, improving generalizability across complex terrain. We use only the bottom 5 kilometers of the profile, giving us a total of 84 input variables ($T$, $T_d$, $U$, $V$ at 21 total height levels).

\subsection{Target Observations}

The mPING effort is a citizen science project developed by the National Severe Storms Laboratory (NSSL) and the University of Oklahoma \citep{MPING}, which collects crowd-sourced weather observations including mixed precipitation types with time and location stamps via GPS. The full archive is freely accessible through a registered API. We chose four targets for classification consistent with most NWP models: rain (RN), snow (SN), freezing rain (FRZR), and ice pellets (IP). Observations of drizzle and freezing drizzle were converted to rain and freezing rain, respectively. 
  
\subsection{Matching Process}
 
Matching of our mPING training targets to the most relevant model profile had multiple phases. First, mixed-type reports were duplicated into one observation per component type — for example, a "rain/snow mix" observation was transformed into one rain observation and one snow observation. Each observation time was then rounded to the following hour and mapped to the nearest 13 km RAP grid cell. For each hourly grid cell containing more than one report, including any duplicated mixed-type components, the most frequent observation type was retained as a single sample. Where frequency ties existed, the observation was selected based on a hazard hierarchy: $RN < SN < IP < FRZR$. This procedure ensures a single observation per space-time location throughout. While duplicating mixed-type observations assumes that a mixed-phase profile is representative of each component type individually, the hourly RAP profiles represent a single atmospheric state that in mixed-type scenarios typically reflects a thermodynamically marginal or transitional environment physically consistent with both component types. The subsequent quality control procedure (Section 3.1) further mitigates this assumption by removing observations whose labeled type is thermodynamically implausible given the matched profile, regardless of their origin. Our matched data spanned 2015-01-01 through 2022-06-30, originally consisting of $\sim$2.55 million samples, which was reduced to $\sim$1.5 million samples following quality control and data curation (Section 3.1). Training data comprised all data before 2020-06-01 with validation data from after this date. The resulting distribution of target classes was: RN: 55\%, SN: 41\%, IP: 2\%, FRZR: 2\%.

\subsection{Scaling}

For ML pre-processing, we re-scale input variables by variable group rather than independently, meaning the scaling parameters are calculated from the combined distribution of all height levels for each of the four variables ($T$, $T_d$, $U$, $V$). The scaling procedure uses the robust scaler method, which subtracts the median and divides by the interquartile range, providing robustness to outliers. Scaling by variable group rather than level ensures that the relative relationships within each variable are preserved across height levels. For example, a temperature at 500m above ground level is interpreted consistently relative to the full temperature distribution rather than relative to the local distribution at that level alone. This approach also improves model interpretability by ensuring that the scaled input space reflects physically meaningful contrasts within each variable. We used the bridgescaler python package, which provides fast group scaling with a scikit-learn-compatible API \citep{bridgescaler}.

\section{Methods}
\subsection{Data Curation and Quality Control}
Initial exploratory analysis of our dataset revealed there was a much wider surface temperature distribution corresponding to p-type events in the mPING data when compared to the p-types output directly by the RAP model. This included freezing precipitation at unphysically warm surface temperatures, liquid rain at extremely cold temperatures, as well as numerous freezing rain and ice pellet reports with unrealistic matching profiles. Examination of collocated radar and atmospheric analysis data revealed that some of the nonphysical mPING reports were not associated with any ongoing precipitation. There are a variety of potential reasons for these inconsistencies: 1) lack of knowledge from the report submitter, 2) adversarial reports \citep{McGovern2019_blackbox}, 3) spatiotemporal mismatches between the analysis data and observations, and 4) atmospheric processes not accurately represented at the vertical resolution of current NWP models. Adversarial reports, for example, could be fabricated to facilitate insurance fraud. Observer knowledge gaps may also play a role. We identified examples where ice pellet reports coincided with summertime convective storms, possibly reflecting confusion between ice pellets and hail given that hail is not listed as a separate option in the mPING app interface. Additionally, the term "sleet" refers to rain/snow mix in parts of the world outside North America \citep{McCabe_2022}, potentially contributing to misclassification by some users. These observational inconsistencies have direct consequences for model training: soundings associated with unfiltered reports produced probabilities of frozen precipitation well above zero even at surface temperatures well above freezing, a behavior that forecasters indicated would engender skepticism and limit trust in operational settings.

\begin{figure*}[b!]
\centerline{\includegraphics[width=\textwidth]{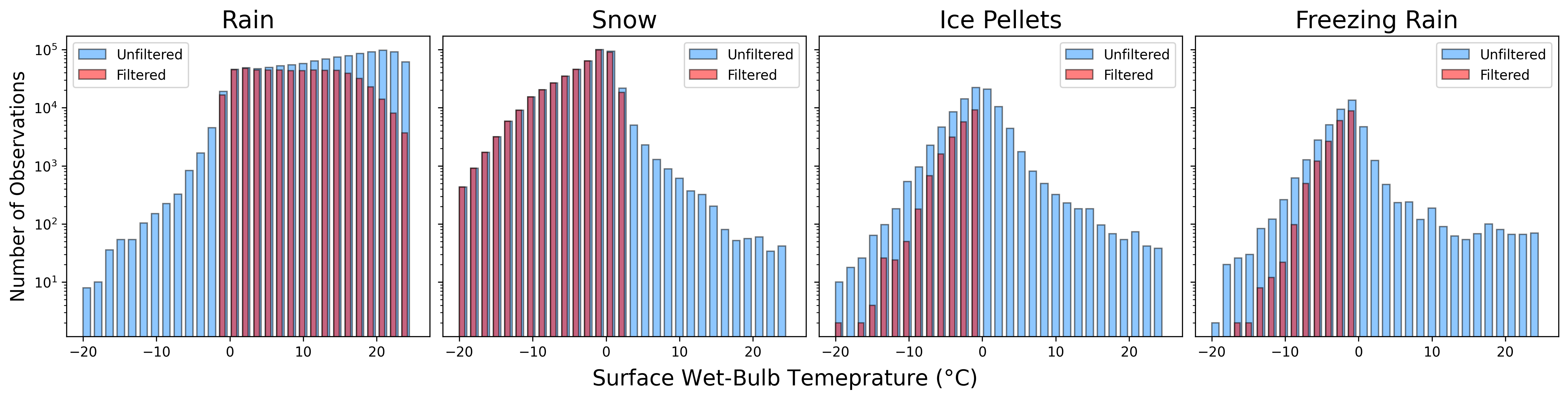}}
\caption{Distributions of mPING observations before (blue) and after (red) quality control was performed.}
\label{fig:ptype_dist}
\end{figure*}

To mitigate the impact of some of these questionable quality reports, we filtered data with mPING labels inconsistent with the wet-bulb temperature at the surface.  Our quality control procedure is listed in Table \ref{tab:qc}. Additionally, we only used observations from October through May to limit the severe class imbalance of precipitation types. This provided us with distributions much more similar to the RAP distributions, albeit still more expansive, though we do not assume the RAP distributions as truth. However, since the RAP model precipitation type is based on the microphysics parameterization, it is likely that they represent realistic environments. See Figure \ref{fig:ptype_dist} for the full mPING distributions before and after quality control. 

\begin{table}[h!]
\centering
\begin{tabular}{|c|c|}
\hline
\textbf{Precipitation Type} & \textbf{Observations removed} \\
\hline
Rain & $T_{\mathrm{wb, sfc}} < -1 ^{\circ}$C \\
\hline
Snow & $T_{\mathrm{wb, sfc}} > 3 ^{\circ}$C \\
\hline
Ice Pellets & $T_{\mathrm{wb, sfc}} > -1^{\circ}$C \\
\hline
Ice Pellets &  $T_{\mathrm{wb}}$ crosses 0~$^\circ$C $< 2$ times \\
\hline
Freezing Rain & $T_{\mathrm{wb, sfc}} > 0 ^{\circ}$C \\
\hline
Freezing Rain & $T_{\mathrm{wb}}$ crosses 0~$^\circ$C $< 2$ times \\
\hline
\end{tabular}
\caption{Quality control procedure for mPING observations.\label{tab:qc}}
\label{tab:my_label}
\end{table}

\subsection{Evidential ML Model Architecture}
Following the evidential deep learning framework for Earth system science described in \citet{evidential-schreck}, we implement an evidential neural network \citep{Sensoy2018-jy} that balances the uncertainty expressiveness of a Bayesian posterior probability distribution with the computational efficiency of a standard neural network. Our evidential p-type model is a multi-layer dense neural network with 4 hidden layers, 200 neurons in each hidden layer, and a Leaky ReLU activation function.
Rather than outputting class probabilities directly, the evidential model outputs positive real values called \textit{evidence}. Higher \textit{relative} evidence for a class boosts its predicted probability, while higher \textit{absolute} evidence concentrates the distribution, indicating stronger overall support. Class probabilities are derived by dividing each class's evidence by the total evidence across all classes. Two forms of uncertainty follow naturally from this framework: aleatoric uncertainty, which reflects irreducible overlap in the training data and is expressed directly through the predicted probability distribution \citep{evidential-schreck}, and epistemic uncertainty, expressed as the inverse of total evidence — a fifth pseudo-class $u$ representing "I don't know." (see the appendix for derivations for epistemic and aleatoric uncertainty from \cite{evidential-schreck}). Higher values of $u$ indicate that the model has accumulated little total support for any class, flagging cases where forecasters should examine contextual information more closely.

\begin{figure*}
\centerline{\includegraphics[width=\textwidth]{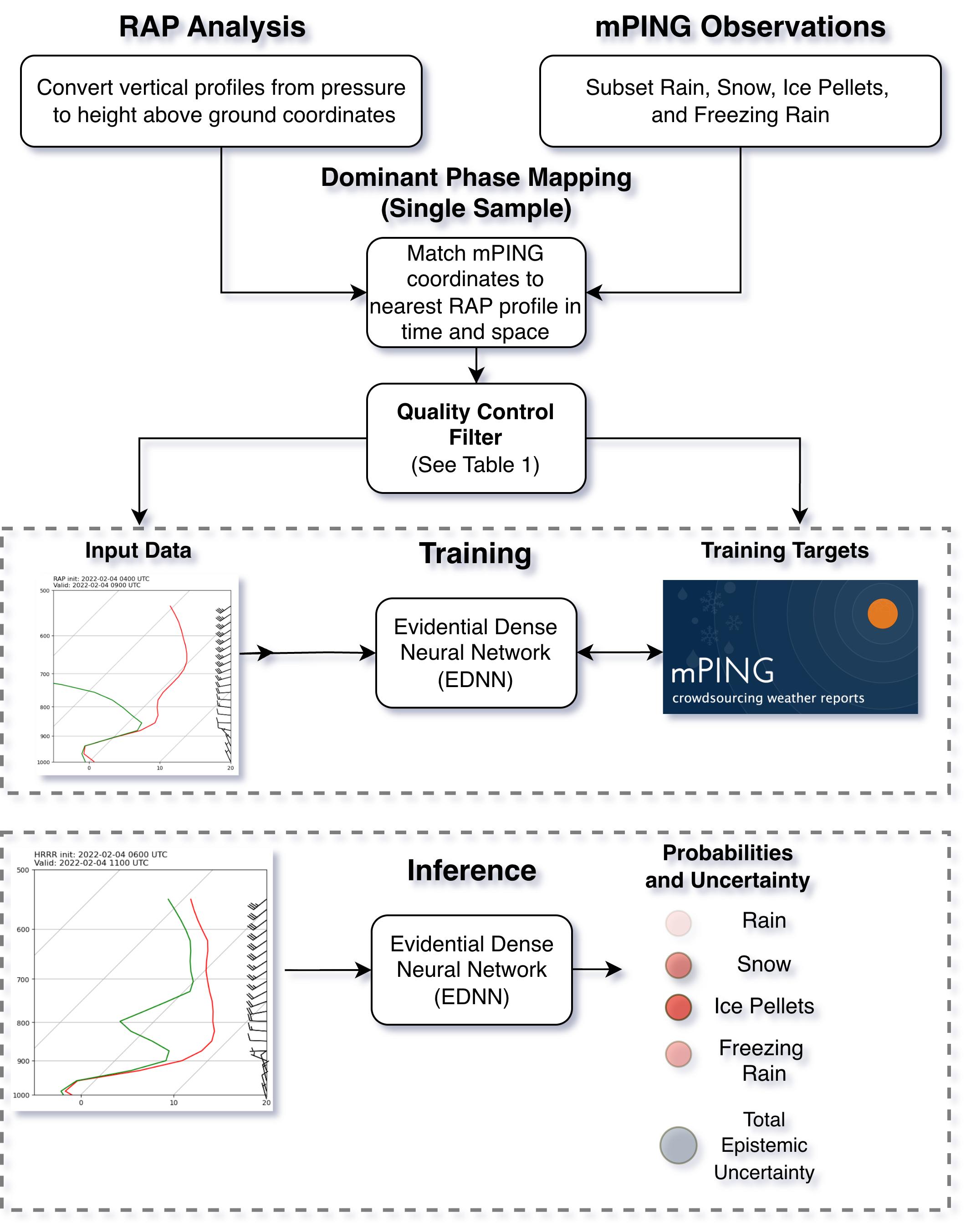}}
\caption{Conceptual diagram outlining the procedure from data origin to model output.}
\label{fig:ptype_concept}
\end{figure*}

The evidential loss function combines two components: an accuracy term that encourages correct class distributions, and a Kullback-Leibler (KL) divergence term that penalizes the model for placing strong evidence on incorrect classes — effectively regularizing the uncertainty estimates and preventing overconfident wrong predictions. The KL term includes a regularization coefficient that requires tuning and can vary considerably across applications \citep{evidential-schreck}. We used the Earth Computing Hyperparameter Optimization (ECHO) framework for distributed hyperparameter searches including optimization of this coefficient (Machine Integration and Learning for Earth Systems 2025). Figure \ref{fig:ptype_concept} provides a methodological overview.

\subsection{Model Comparison}

The original and modified Bourgouin methods use the integrated thermal energy on both sides of the 0°C isotherm (constrained to below the point in the atmosphere where the dry-bulb or wet-bulb profile first crosses the freezing line in its descent) to heuristically determine precipitation type based on a curve fit to observations in conjunction with heuristics. Observations falling above the decision curve are classified as ice pellets and those below as freezing rain. In the original method, energy is calculated using the dry-bulb temperature; the modified version \citep{Birk2021-ch} uses the wet-bulb temperature, which accounts for evaporative cooling and generally produces more physically consistent results. We follow \cite{Birk2021-ch} in converting the inherently negative freeze energy to a positive value for comparison, and use wet-bulb temperature profiles consistently throughout our thermodynamic energy analysis.

The Modified Bourgouin technique utilizes marginal probabilities derived from independent thermodynamic thresholds, operating effectively as a multi-label system where the sum of probabilities is not constrained to 1.0. This independent structure naturally accommodates the occurrence of mixed precipitation. In contrast, the ML model outputs joint probabilities normalized to 1.0, treating the primary precipitation types as mutually exclusive classes and requiring the analysis of secondary probabilities to potentially diagnose phase mixtures or uncertainty. 

Evaluation of all methods presents challenges due to biased or sparse ground truth data. We compare the ML model against held-out mPING observations, the deterministic RAP and High Resolution Rapid Refresh (HRRR) NWP models, and both the original and modified Bourgouin area methods. Since the area methods and our ML approach are both conditioned on precipitation existing, all evaluation data are subset to grid points where the NWP model indicates non-zero precipitation. The RAP and HRRR have separate binary diagnostics for each precipitation type and can represent mixed types simultaneously; for these cases the same hazard hierarchy used during training ($RN < SN < IP < FRZR$) was applied to produce a single dominant prediction.

\section{Results}
\subsection{Bulk Statistics}

\begin{figure}[t!]
\centering
 \noindent\includegraphics[width=\textwidth,angle=0]{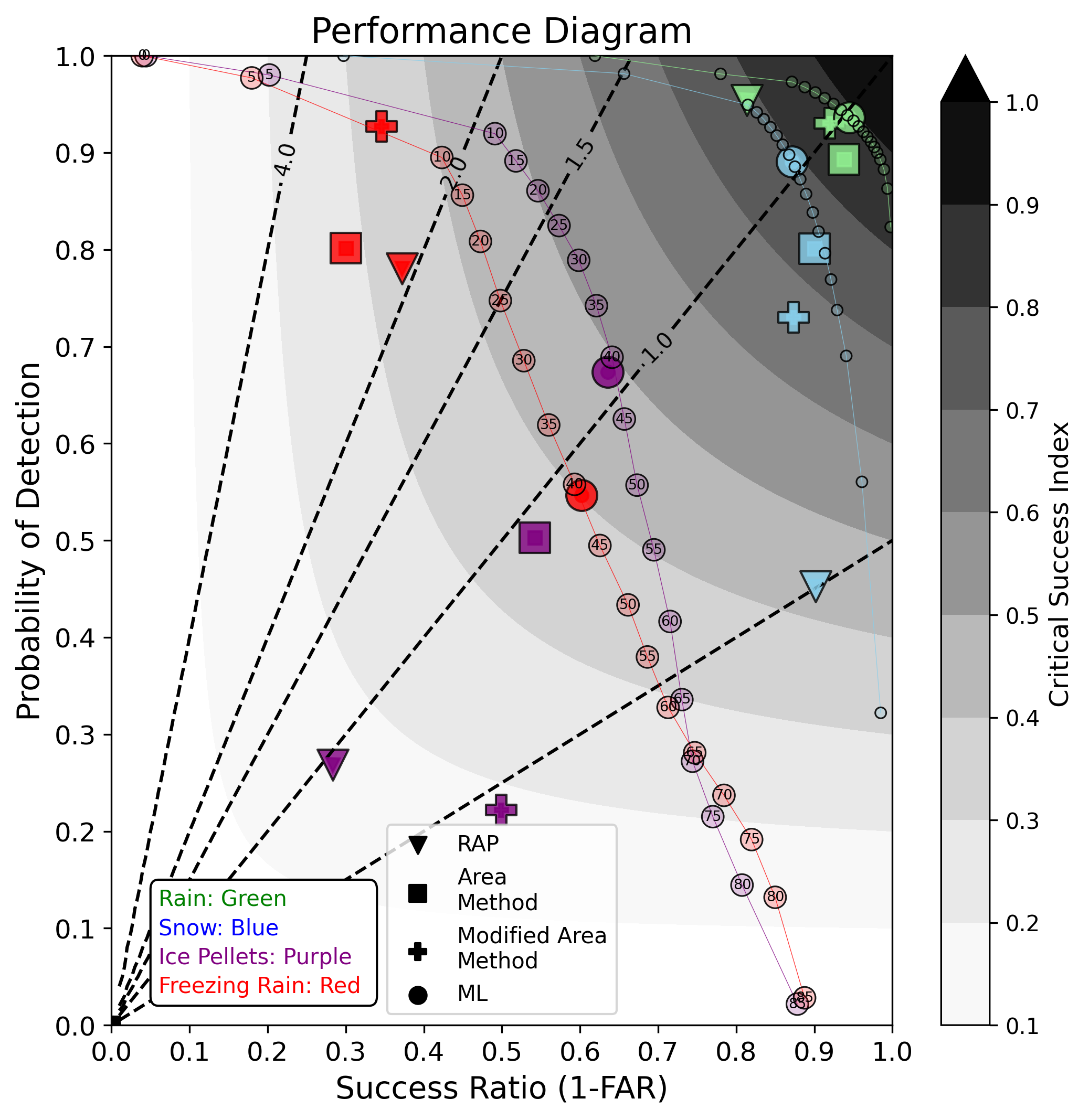}\\
 \caption{Bulk Performance metrics for all models by precipitation type. The large markers use the dominant class. The shaded lines represent the ML model performance with probability thresholds decreasing from 1.0 to 0.0 every 0.05 (left to right).}\label{fig:perf}
\end{figure}

Figure \ref{fig:perf} demonstrates the performance of all four modeling methods highlighting the probability of detection (POD): $\frac{hits}{hits \ + \ misses}$ on the y-axis, success ratio: $\frac{hits}{hits \ + \ false \ alarms}$ on the x-axis, and the critical success index: $\frac{hits}{hits \ + \ misses \  + \ false \ alarms}$ in the shaded regions for all 4 classes. The large markers represent the bulk metric using the dominant predicted p-type. The shaded curves represent the ML model's performance when thresholds are applied by probability (with decreasing 0.05 probability thresholds from left to right). 

Notably, when looking at the bulk metrics, all non-ML methods have a significantly higher probability of detection for freezing rain than the ML method, but many more false alarms as evident by much lower success ratios. The modified area method in particular has a staggeringly high bulk POD. The ML model needs a low threshold of 0.21 probability to recover an equal POD as the original Bourgouin method, and a threshold of 0.08 to recover the POD of the modified area method. For ice pellets, the ML model improves significantly for all metrics compared to the modified Bourgouin method.  For the remaining two classes, the ML approach generally performs better or comparably. 

\begin{figure}[!htt]
\centering
 \noindent\includegraphics[width=\textwidth,angle=0]{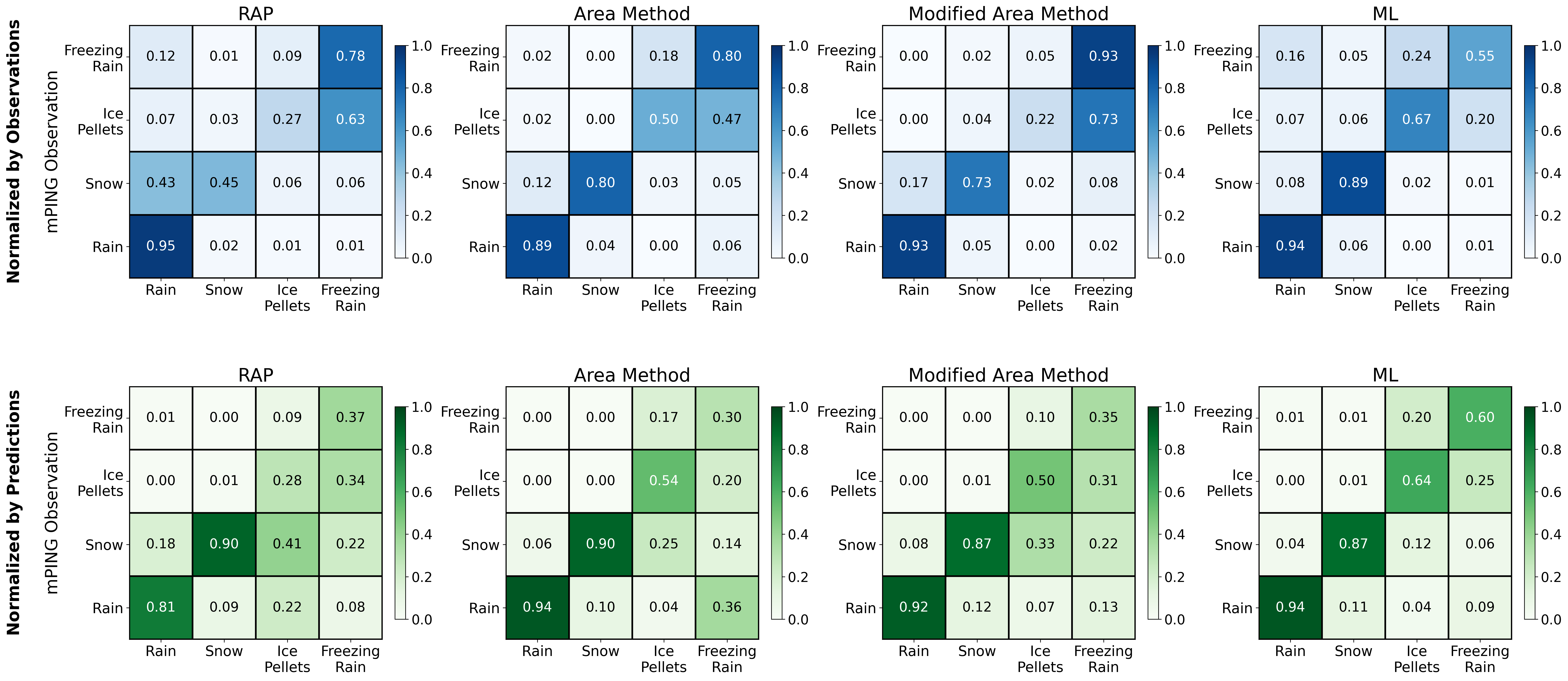}\\
 \caption{Contingency tables for all models. The top row of table is normalized by the mPING observations (rows), and the bottom row is normalized by model predictions (columns). The predicted class is represented on the x-axis and the observations on the y-axis. }\label{fig:cm}
\end{figure}

To evaluate the model's categorical performance, a multi-class contingency table (Figure \ref{fig:cm}) cross-tabulates predicted precipitation types against actual surface observations. 
Normalizing by total observations (top row) yields the Probability of Detection (POD) on the diagonal, while the off-diagonals explicitly detail the distribution of misses—showing which categories the model erroneously forecasted when a specific event was occurring. Conversely, normalizing by total predictions yields the Success Ratio (1 - False Alarm Ratio) on the diagonal (bottom row), with the off-diagonals revealing the source of false alarms by showing what was actually occurring when the model incorrectly predicted a given category.

In the top row, we confirm that the ML model had the lowest probability of detection for freezing rain, and that the second and third most common predictions when freezing rain is observed are ice pellets and rain, respectively.  This distinction between rain and freezing rain, such as when the vertical profile has an elevated warm layer and a surface freezing layer that is not strong enough to supercool the droplets, is something that the Bourgouin methods cannot account for, and is better represented by the ML and RAP models. The bottom row shows that the success ratio is significantly higher for ice pellets and freezing rain with the ML model than all other methods. The non-ML methods tend to often predict ice pellets in a snow regime, and predict freezing rain in many scenarios. The ML method appears to constrain its false alarms mostly between freezing rain and ice pellets in those specific cases. 
 
\begin{figure}[h!]
\centering
 \noindent\includegraphics[width=\textwidth,angle=0]{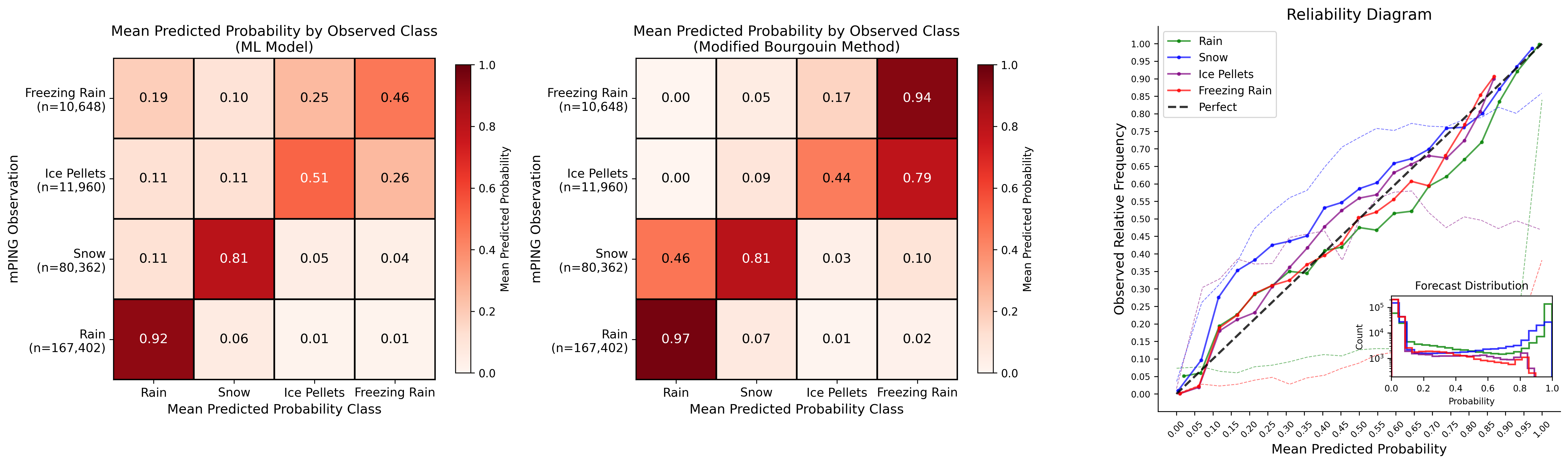}\\
 \caption{A mean predicted probability matrix for the ML model (left) and modified area method (center). Reliability diagrams for the ML model (dark solid) and modified area method (shaded dashed) for each p-type (right).}
 \label{fig:reliability}
\end{figure}

While categorical contingency tables evaluate discrete outcomes, they inherently mask a model’s underlying certainty. To assess probabilistic skill, we utilize a mean probability matrix paired with a reliability diagram for both probabilistic models (Figure \ref{fig:reliability}). The mean predicted probability matrix averages the predicted probabilities assigned to each precipitation category during specific observed events. This averaging reveals how the model allocates its confidence in ambiguous environments, highlighting whether it captures appropriate uncertainty or suffers from being confidently wrong. The reliability diagram then evaluates forecast calibration by plotting those predicted probabilities against observed relative frequencies. Well-calibrated forecasts track the diagonal, while deviations diagnose over-confidence (points below the line) or under-confidence (points above). 

Similar to the bulk tables, the mean probabilities for the ML model (left) are lower for freezing rain (0.46) and ice pellets (0.51), and the majority mean misses for each are concentrated in the other. Also notable is the relatively high probability of 0.19 for a rain prediction with a freezing rain observation. For the modified area method (center), we see a very high mean freezing rain probability for observed freezing rain events. Also notable is the lower mean ice pellet verification paired with very high freezing rain probabilities and a similar trend with snow verification and rain which demonstrates the usefulness of the marginal probability approach to potentially capture mixed events and/or uncertainty. On the right, we see bulk reliability diagrams for the ML model (dark solid lines) and modified area method (light dashed lines). Although the ML model is relatively well-calibrated, it demonstrates slight over-forecasting of rain, ice pellets, freezing rain at higher probabilities, and an underforecasting of those types at lower probabilities. Snow is notably under-forecasted at lower probabilities. The modified area method is very poorly calibrated with extremely high over-forecasts of freezing rain and rain, and severe underforecasts of snow which helps explain the very high POD and false alarms for freezing rain.  

\begin{figure}[h!]
\centering \noindent\includegraphics[width=\textwidth, angle=0]{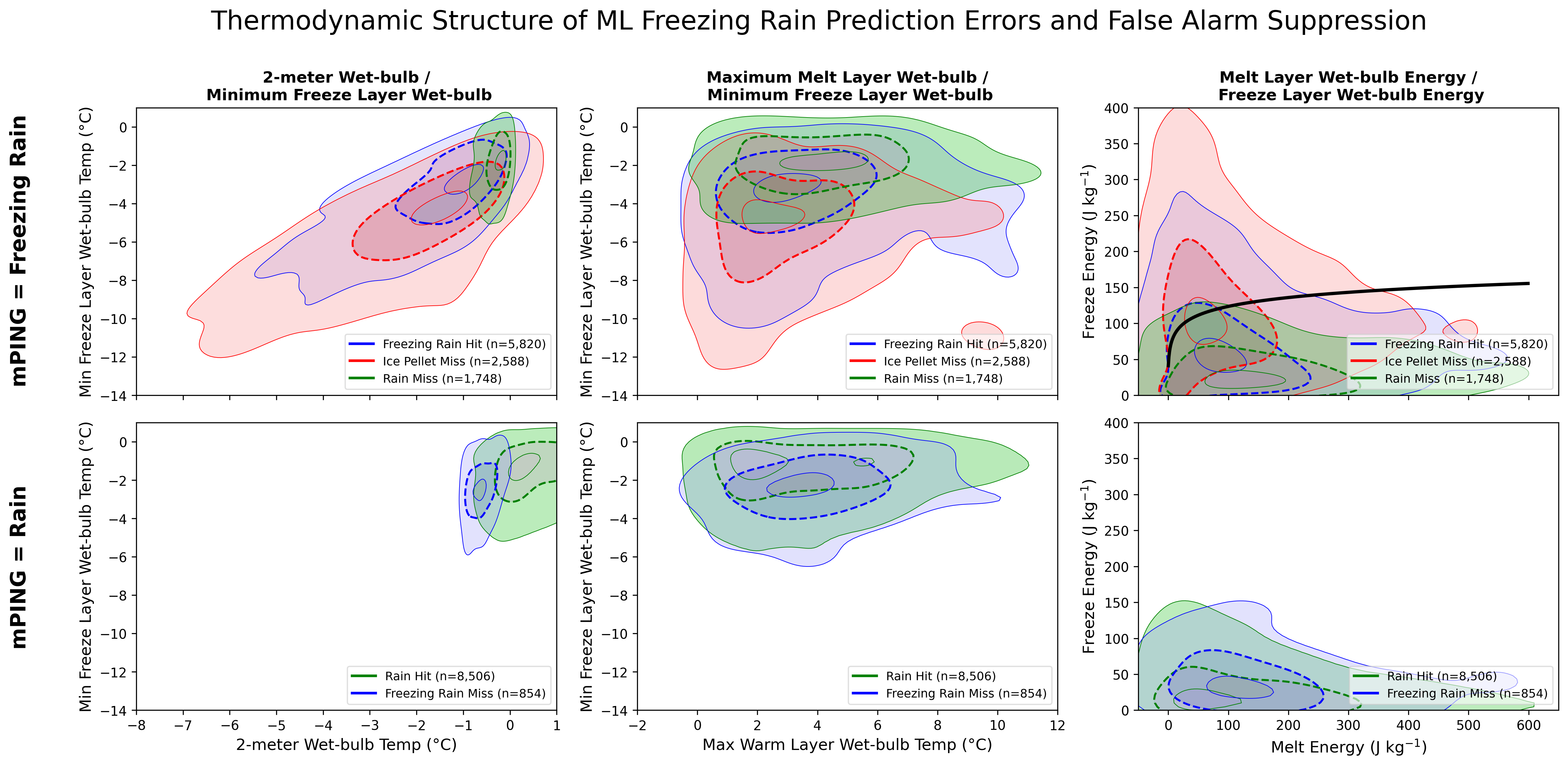}\\
 \caption{Kernel density estimates of hits and misses in 2D diagnostic space conditioned on observation type. The top row is conditioned on freezing rain observations and the bottom conditioned on rain observations. The modified Bourgouin curve for ice pellet and freezing rain classification is overlaid in black. }\label{fig:uncertain}
\end{figure}

Figure \ref{fig:uncertain} shows kernel density estimates of ML prediction categories for mPING-observed freezing rain (top row) and mPING-observed rain in near-freezing environments (surface wet-bulb temperature below 2°C with a detectable freeze layer) across three thermodynamic diagnostic spaces: surface wet-bulb temperature versus minimum freeze layer wet-bulb temperature (left), minimum freeze layer temperature versus lower 0°C crossing height (center), and melt versus freeze energy in the Bourgouin framework (right). Three contours for each type are represented which enclose enclose 90\%, 50\%, and 10\% of each category's distribution respectively. In the top row, correct freezing rain predictions (blue), ice pellet misses (red), and rain misses (green) show clear separation at the 90\% density contour across all three diagnostic spaces, while approximately 50\% overlap exists at the 50\% contour, indicating that the typical case for each prediction category occupies a physically distinct thermodynamic regime, but that genuine atmospheric ambiguity persists.

The thermodynamic structure of ML prediction errors is physically consistent across all three diagnostic spaces (Fig. \ref{fig:uncertain}). Ice pellet misses (red) occur systematically in environments with colder freeze layers and colder max warm layer wet-bulb temperatures — thermodynamic conditions more characteristic of ice pellet formation where greater fall distance and more intense refreezing potential are present. Rain misses (green, top row) occur in environments with warmer freeze layers and warmer surface wet-bulb temperatures, occupying the same thermodynamic corner as correct rain predictions in near-freezing environments (green, bottom row). This spatial alignment across rows demonstrates that the ML model consistently treats this warm, shallow freeze layer regime as rain-like regardless of the observed precipitation type, reflecting physically coherent behavior at the FZRA/RN thermodynamic boundary.

Cases where the ML model predicts freezing rain but mPING reports rain (blue, bottom row) are thermodynamically distinct from correct rain predictions (green, bottom row) but closely aligned with correct freezing rain predictions (blue, top row), particularly in surface wet-bulb temperature and freeze layer intensity space. The separation between them is most pronounced in the left panel, where freezing rain misses are systematically colder at the surface by approximately one degree, and least pronounced in the Bourgouin energy space (right panel) where the two groups show significant overlap at the 50\% density contour. This progressive loss of discriminability from left to right across the diagnostic spaces suggests that the ML model has learned surface-based relationships not captured by the modified area method, providing meaningful additional information for distinguishing rain-like from freezing rain-like environments near the thermodynamic boundary.
  
\begin{figure}[h!]
\centering \noindent\includegraphics[width=\textwidth, angle=0]{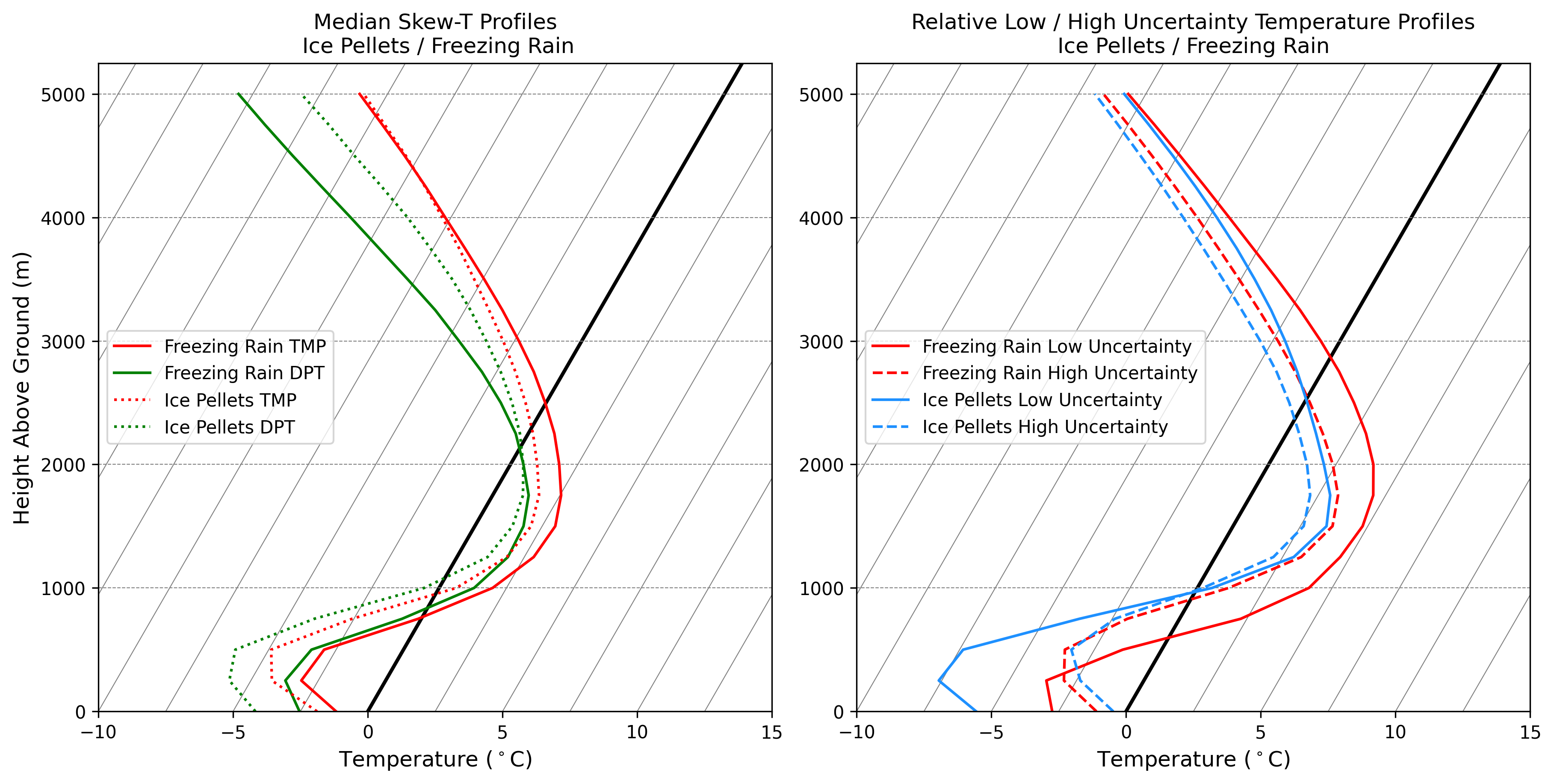}\\
 \caption{Composite profiles for ice pellets and freezing rain for entire validation set (left). The right panel are the composites of the low ($<$ 0.05) and high ($>$ 0.95) quantiles of epistemic uncertainty.}\label{fig:icep_frzr}
\end{figure}

To further examine the thermodynamic basis of the ice pellet and freezing rain miss regimes identified above, Figure \ref{fig:icep_frzr} shows median composite profiles for correct ML predictions of each type across the full validation set. The ice pellet profile exhibits a lower melt energy and higher freeze energy in addition to a drier surface freeze layer, which would increase evaporative cooling and support refreezing at a higher rate. The right panel showcases the respective high and low uncertainty soundings (epistemic uncertainty quantiles of 0.05 and 0.95). The low uncertainty cases appear to be clear examples of each type, while the high uncertainty cases are nearly identical, highlighting the model's ability to express genuine uncertainty and the potential limits of predictability in these regimes.

\subsection{Case Studies}

To complement the aggregate statistical evaluation, we examine two case studies that illustrate how the ML model's probabilistic output and physical consistency manifest across contrasting operational scenarios. We verify our model over two distinct and difficult thermodynamic setups for precipitation type forecasting winter storms: the Christmas 2023 winter storm centered around Aberdeen, SD (KABR), and the January 22–24, 2024, freezing rain event near Champaign, IL (KCMI). The KABR event was a deep-synoptic system featuring rapid spatial transitions between rain, severe ice accumulation, and blizzard conditions. In contrast, the January 22-24 event represents a classic, prolonged freezing rain setup driven by powerful warm air advection overrunning an extremely shallow sub-freezing arctic surface layer. Together, these cases rigorously evaluate the model’s physical consistency across the two difficult forecasting regimes: rapidly evolving synoptic transition zones and persistent, shallow cold domes. Due to the lack of spatial and temporal consistency of mPING observations during these storms, we chose to use the KABR (Aberdeen, SD) and KCMI (Champaign, IL) ASOS stations for verification to analyze a time series of observations despite the ASOS stations' inability to detect ice pellets.

\subsubsection{Christmas 2023 Winter Storm}

\begin{figure}[h!]
\centering \noindent\includegraphics[width=\textwidth, angle=0]{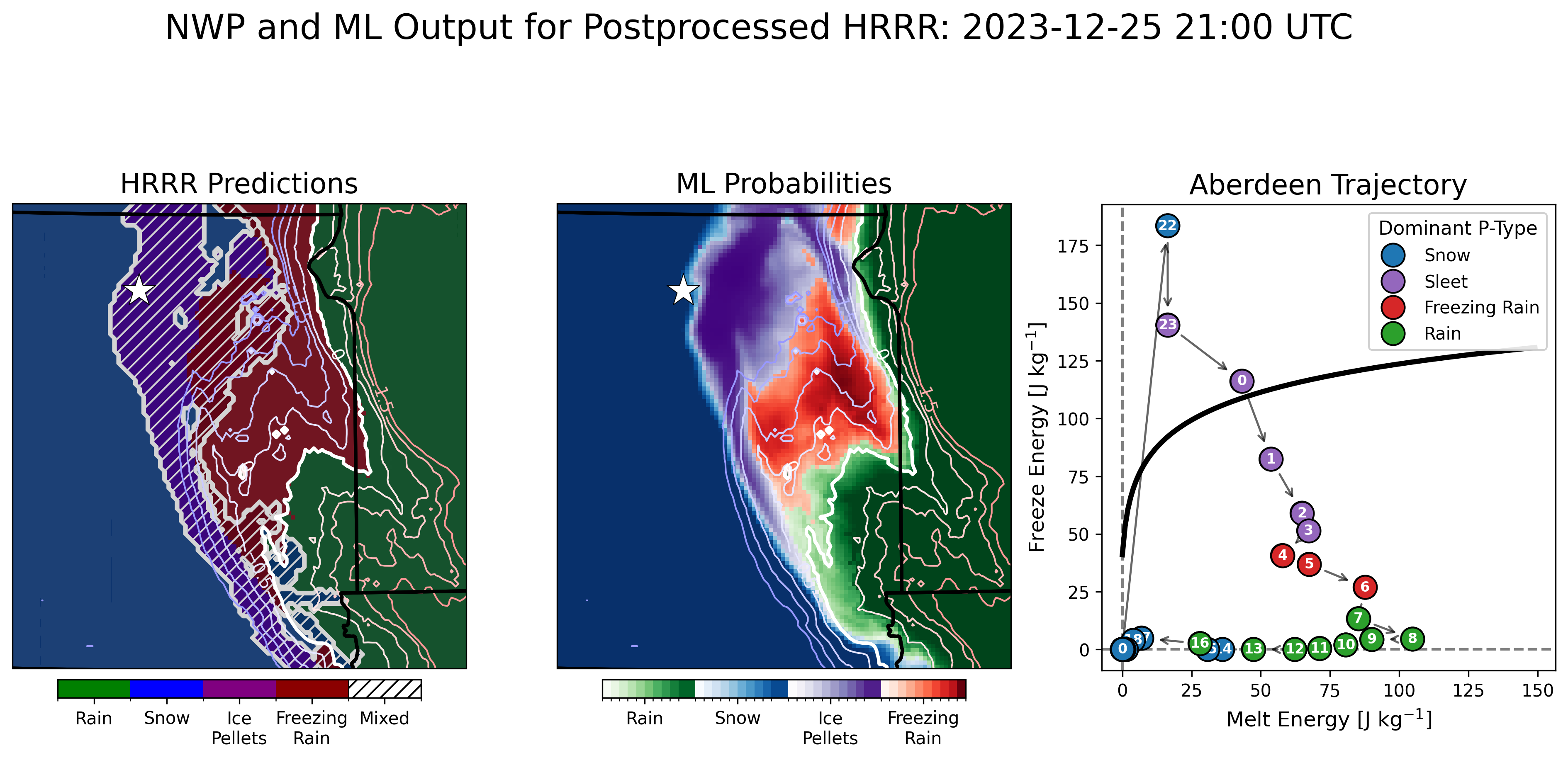}\\
 \caption{HRRR predictions by class (left), ML probabilities by predominant class (center, shaded by probability), and trajectory of dominant phase ML predictions in wet-bulb energy space at Aberdeen, SD. The white contour is the  0 $^{\circ}$C isotherm with blue and red contours representing the surface temperature at -/+ 0.5 $^{\circ}$C intervals. The white star is the Aberdeen ASOS station location. }\label{fig:spatial}
\end{figure}

Figure \ref{fig:spatial}'s left and center panels provide a snapshot of the system at 21:00Z on 2023-12-25. The right panel shows the ML prediction trajectory at the ASOS site in Aberdeen, SD (marked on the other panels with a white star) in Bourgouin wet-bulb energy space spanning 2023-12-25 00:00Z to 2023-12-27 00:00Z. The maps highlight the transition region covering all four p-types. The left panel (HRRR) reports mixed types in the freezing rain and ice pellet favorable environments as well as the western region where ice pellets transition to snow. The HRRR predicts mixed snow and ice pellets in a longer southern swath compared with the ML model (center), which is much more confident in snow in that region. The central mixed transition between ice pellets and freezing rain corresponds to the lower probabilities in the ML model in a highly ambiguous thermodynamic region. Notably, the transition of the HRRR model from rain to freezing rain corresponds directly to the 2-meter surface temperature 0 $^{\circ}$C isotherm (white solid line). The ML method highlights the latter with low probabilities of rain and freezing rain in regions where the near-surface temperature roughly ranges from $-1.0 \,^{\circ}\text{C} \text{ to } 0.5 \,^{\circ}\text{C}$ and highlights the ML model's learning of thermodynamic space that is not overly conditioned on near-zero degree surface temperatures and can reflect uncertainty in complex spatial transition zones.  

In the Figure \ref{fig:spatial}'s right panel, we see the point trajectory with the dominant predicted type from the ML model in wet-bulb energy space. The modified Bourgouin method curve for discriminating between ice pellets and freezing rain is overlaid in black. Generally, the evolution is energy space appears physically consistent with transitions from snow to ice pellets with increasing melt energies where snow would be unlikely to survive the warm descent. As the evolution continues, we see a shift from ice pellets to freezing rain, and then rain, with freeze layer energies below $50 \, \mathrm{J \, kg}^{-1}$. Lastly, with very small amounts ($< 10 \, \mathrm{J \, kg}^{-1}$) freeze energy and decreasing melt energy, we see the transition from rain to snow. There are both some similarities and differences when comparing to the \cite{Birk2021-ch} area method. Similarly, the ability of snow to survive a melting layer with melt energies near $30 \, \mathrm{J \, kg}^{-1}$ is consistent in both high and low freeze energy predictions. The main difference is the ML models favoring of ice pellets instead of freezing rain, especially at 2023-12-26 02:00Z and 2023-12-26 03:00Z, which are squarely in the "freezing rain dominant" zone of the modified method which would result in high probabilities of freezing rain. 

 \begin{figure}[b!]
\centering
\noindent\includegraphics[width=\textwidth,angle=0]{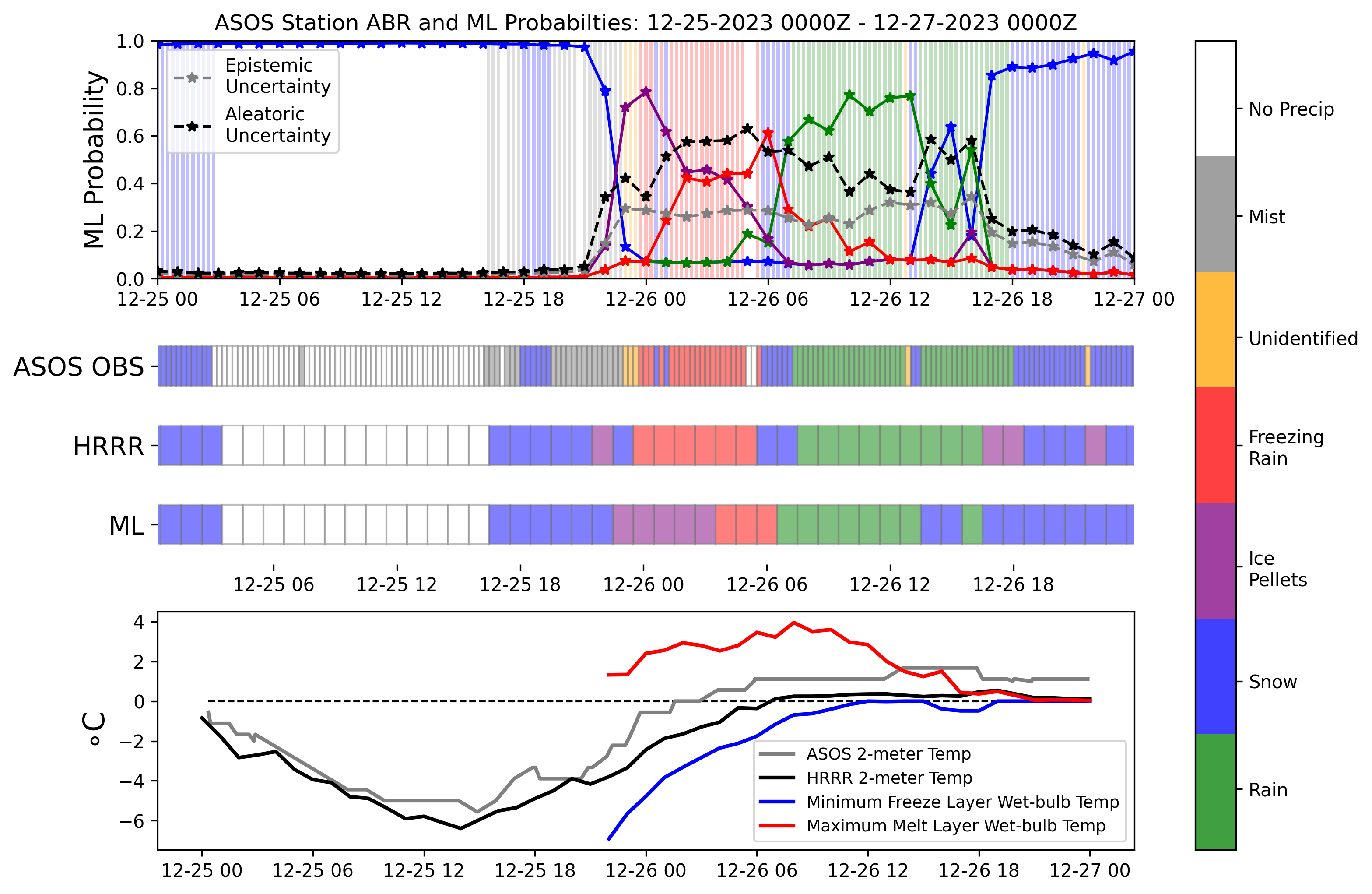}
 \caption{Time series from the Aberdeen, SD ASOS station with probabilities and uncertainties from the ML model post processed from HRRR analysis data. The lightly shaded lines in the top row are 15-minute ASOS observations. The center panel represents the class observations and predictions. The bottom row shows the observational and analysis surface temperatures as well as the minimum freeze layer and maximum melt layer wet-bulb temperatures.}
 \label{fig:time}
\end{figure} 

Figure \ref{fig:time} is a time series analysis at the Aberdeen, South Dakota, ASOS station (ABD) in which all four p-types were likely observed during a 12-hour period. The upper panel shows the ML probabilities and uncertainties derived from HRRR analysis along with the observed ASOS conditions. The center panel shows ASOS observations and the prediction type for both the HRRR and ML models.  In general, both the HRRR and ML approaches match up fairly well with the observations, but there are some notable differences. The ML model appears to favor ice pellets when discriminating between ice pellets and freezing rain from 00Z-06Z on 26 Dec, though the probabilities for each are quite similar and it exhibits very high epistemic uncertainty. Secondly, the ML model seems to capture the transition to snow at 18Z on 26 Dec, with a relatively high probability, much better than the HRRR  which transitions from rain to ice pellets before transitioning to snow. The thermodynamic uncertainty is represented in the probability evolution, notably, with the low probabilities (and high derived aleatoric uncertainty) in the 18-hour window starting at 2023-12-25 22:00Z where we see numerous transitions. We can see the two ice pellet predictions from figure \ref{fig:spatial} having near identical probabilities to that of freezing rain. The probability lines also show useful uncertainty information regarding which classes the model is struggling to capture, with clear uncertainty showing up between IP / FRZR, RA / FRZR, and RA / SN.  The bottom panel also helps ground this uncertainty in difficult scenarios with the observation and analysis surface temperatures plotted along with the maximum magnitudes of the melt and freeze layer wet-bulb temperatures. The ASOS observations also reveal some interesting behavior. First, we see that from approximately 2023-12-26 00:04Z to 2023-12-26 06:00Z we see that the observations record non-freezing 2-meter temperatures, upwards of 1 $^{\circ}$C  and yet still observe freezing rain. Secondly, ASOS instrumentation does not currently have the capacity to detect ice pellets, but it does have an "unidentified" p-type, which may correspond to ice pellets or other mixed transition zones.  

\begin{figure}[!b]
\centering

 \noindent\includegraphics[width=36pc,angle=0]{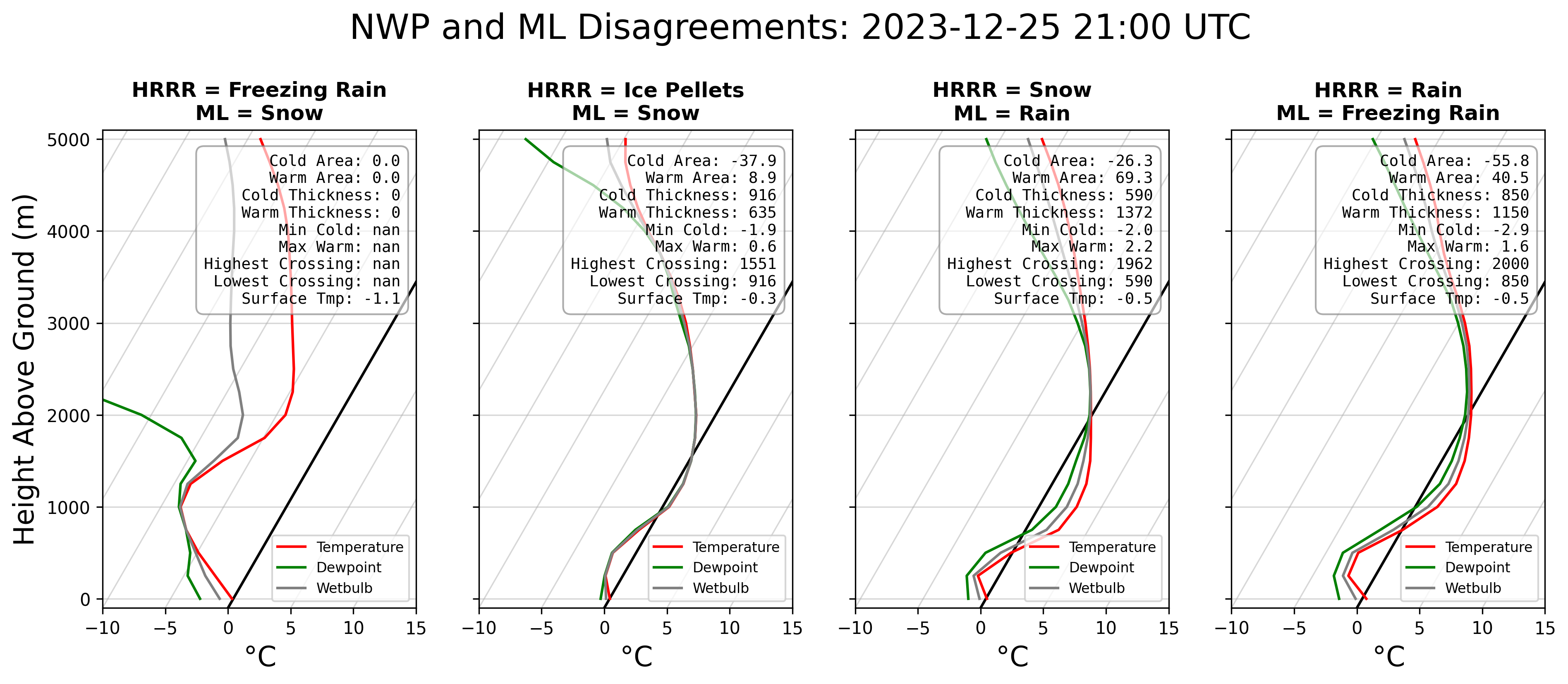}\\
 \caption{Vertical profiles for 2023-12-25 2100 UTC sample where a HRRR prediction (non-mixed) disagreed with the ML prediction.}\label{fig:disagree}
\end{figure}

Lastly, we inspect a few instances where there was disagreement between the models from various parts of the domain at the same time 2023-12-25 2100 UTC. For simplification, we exclude any examples containing HRRR mixed types. Example 1 (left panel) of Figure \ref{fig:disagree} is perhaps the most striking in that the HRRR model predicted freezing rain with the complete absence of an elevated melting layer. This is perhaps possible due to the combination of the microphysical scheme and heuristic thresholds to increase the overall probability of detection for freezing rain and ice pellets \citep{Benjamin2016-xz, Manikin_2005}. Panel 2 is a case where a very small elevated melting layer may or may not be enough energy to melt a snowflake. Panel 3 has a high melt energy, though mid level saturation and snowfall rate may be responsible for the HRRR microphysical scheme to not fully melt the snow. Panel 4 is a classic freezing rain profile until it reaches the surface where the temperature is just above freezing in which the HRRR model heuristically chooses rain instead of freezing rain. This is clearly demonstrated in figure \ref{fig:spatial} where the boundary between rain and freezing rain is directly aligned with the 0 $^{\circ}$C isotherm. 

\subsubsection{2024-01-22 - 2024-01-24 Freezing Rain Event}

\begin{figure}[!b]
\centering

 \noindent\includegraphics[width=36pc,angle=0]{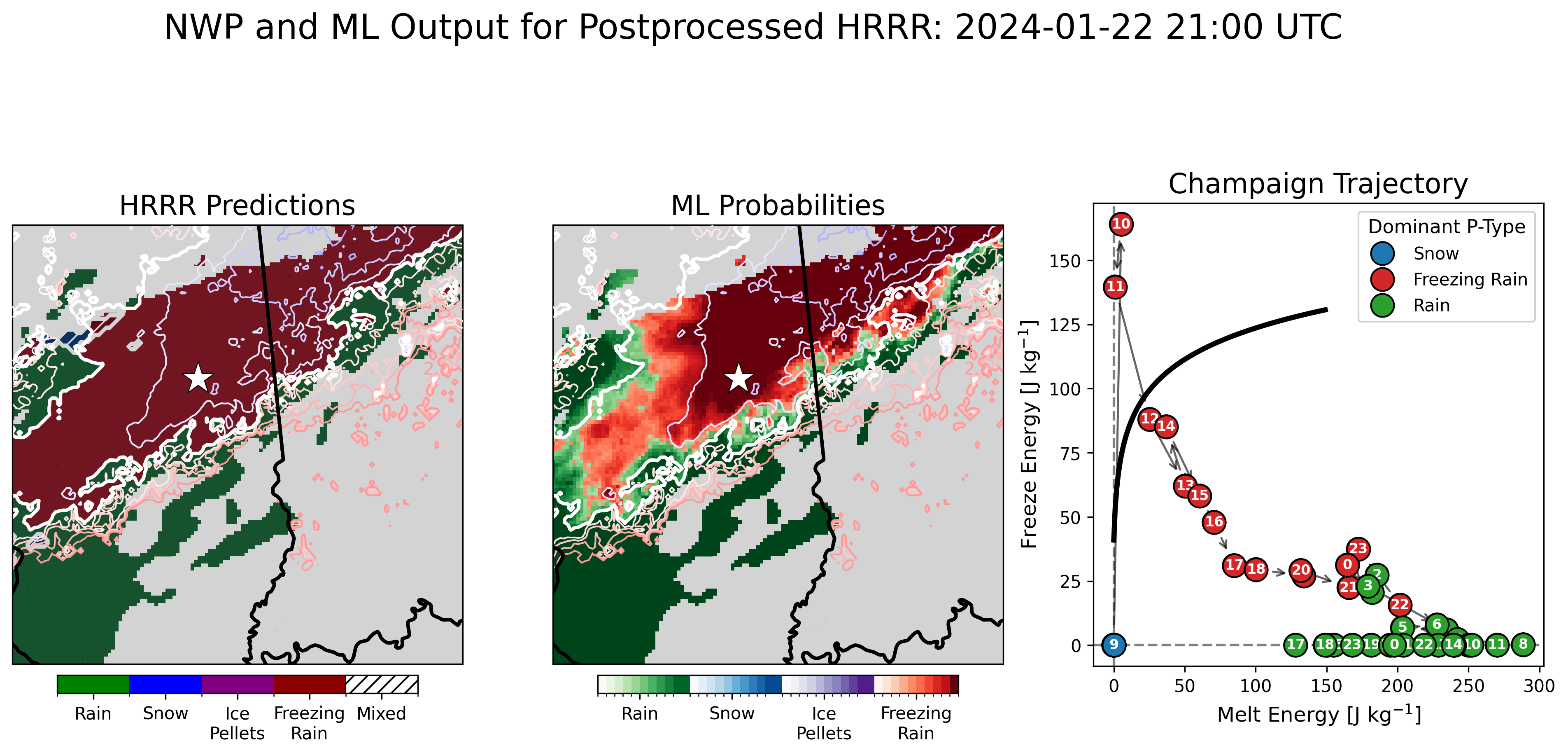}
 \caption{HRRR predictions by class (left), ML probabilities by predominant class (center, shaded by probability), and trajectory of dominant phase ML predictions in wet-bulb energy space at Champaign, IL. The white contour is the 0 $^{\circ}$C isotherm with blue and red contours representing the surface temperature at -/+ 0.5 $^{\circ}$C intervals. The white star is the Champaign ASOS station location.}\label{fig:spatial_KCMI}
\end{figure}
\begin{figure}[!t]
\centering

 \noindent\includegraphics[width=36pc,angle=0]{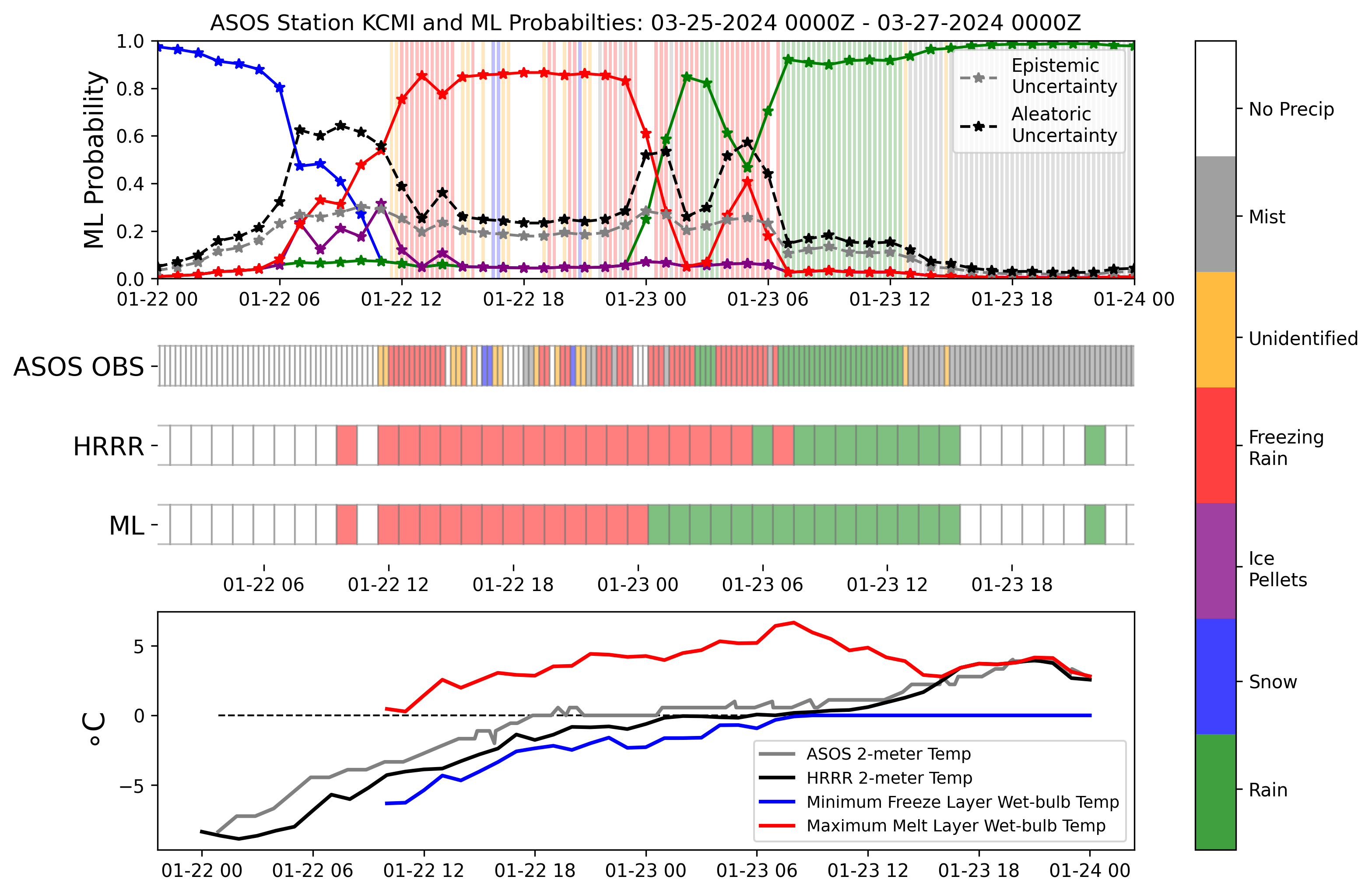}
 \caption{Time series from the Champaign, IL ASOS station with probabilities and uncertainties from the ML model post processed from HRRR analysis data. The lightly shaded lines in the top row are 15-minute ASOS observations. The center panel represents the class observations and predictions. The bottom row shows the observational and analysis surface temperatures as well as the minimum freeze layer and maximum melt layer wet-bulb temperatures.}\label{fig:temporal_KCMI}
\end{figure}

The January 22–24, 2024 event presented a contrasting thermodynamic scenario to the Christmas 2023 storm: a prolonged freezing rain setup driven by strong warm air advection overrunning a shallow sub-freezing surface layer, with little involvement of ice pellets or snow (Figures \ref{fig:spatial_KCMI} and \ref{fig:temporal_KCMI}). One striking feature is the strong and near immediate transition of freezing rain probabilities from high to low probability (with it still being the dominant type) and both covering a large spatial area. We also see an even larger region of dominant rain predictions with sub-zero 2-meter wet-bulb temperatures compared to the previous study. In the trajectory panel, we also see an interesting prediction at 2024-01-23 00:00 that exhibits higher melt energy and lower freeze energy than several rain predictions, suggesting the ML model has learned additional discriminating relationships beyond those captured by the Bourgouin decision boundary. 

The findings in figure \ref{fig:temporal_KCMI} demonstrate a scenario where the ML model has very high probabilities for freezing rain from 2024-01-22 13:00Z to 2024-01-22 23:00Z. ASOS observations during this window are predominantly freezing rain but include intermittent reports of mist, no precipitation, snow, and unidentified types, reflecting the complex and transitional nature of the event. Observations show a brief 1-hour transition to rain and then back to freezing rain at 2024-01-23 04:00Z and the ML model reflects this with a significant increase in rain probability. However, the ML model predicts rain in the hours immediately before and after this transition despite ASOS-observed freezing rain, though with notably higher uncertainty than during the core freezing rain period.

\subsection{Interactive Analysis}

\begin{figure}[h]
\centering
 \noindent\includegraphics[width=\textwidth, angle=0]{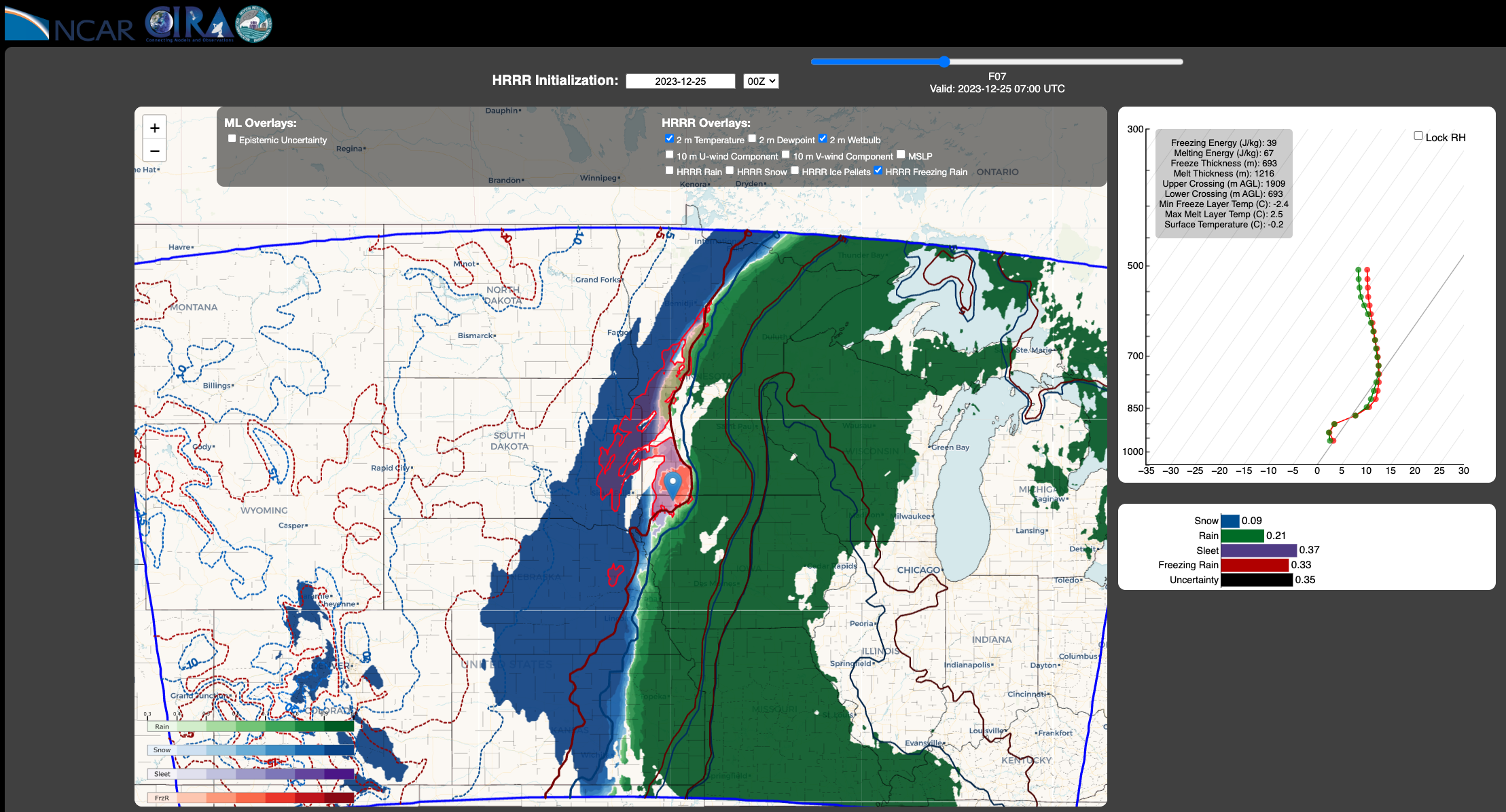}\\
 \caption{A screenshot from the interactive web viewer during the Christmas 2023 winter storm. The sounding on the right is the preprocessed HRRR forecast for input into the model, along with computed diagnostics above. The models probabilities and epistemic uncertainty are shown beneath it. Users have the ability to click and drag the sounding which is dynamically fed into the model to interrogate model output in real-time. }\label{fig:interactive}
\end{figure}

Accurately conveying model predictions and uncertainties through interactive tools can increase forecaster trust and provide a framework for direct feedback \citep{Novak2023-th, rogers2023nws}. To address this need, we developed an interactive web application for visualizing near real-time or historical ML model output that supports real-time dynamic operational interrogation of model performance and physical interpretability (Figure \ref{fig:interactive}). The application allows spatial navigation and time scrolling of model output with overlay of relevant meteorological features, and provides instantaneous display of all p-type probabilities and epistemic uncertainty at any location. Users can retrieve the NWP vertical profile used as model input at any point, with derived thermodynamic statistics displayed alongside the sounding and the model probabilities and uncertainty directly below. The sounding profile is itself interactive: users can modify it by clicking and dragging to create arbitrary profiles, with model predictions updating in real time. This dynamic linking between user-controlled profile perturbations and model inference enables direct testing of physical consistency and diagnosis of failure modes, including out-of-distribution inputs that aggregate verification statistics cannot reveal. For example, the disagreement profiles shown in Figure \ref{fig:disagree} illustrate cases where the interactive tool could be used to explore how small modifications to the warm nose or freeze layer, such as those that might arise from NWP analysis errors, would affect model predictions, providing forecasters with direct insight into the sensitivity of predictions to thermodynamic uncertainty.

\section{Discussion}

The results presented here demonstrate that an evidential neural network trained on raw vertical thermodynamic profiles can produce physically consistent, well-calibrated probabilistic winter precipitation type forecasts that compare favorably to both deterministic NWP-based methods and established thermodynamic diagnostic methods. The reduced freezing rain probability of detection, discussed in detail below, reflects physically structured miss regimes in genuinely ambiguous thermodynamic environments rather than a uniform model deficiency, and is more honestly represented through the full probability distribution than through the dominant predicted class alone. The following discussion examines the physical basis for the ML model's performance characteristics and situates these findings within the broader context of observational uncertainty, physical interpretability, and uncertainty quantification.

\subsection{Thermodynamic Regime Analysis and Probabilistic Consistency}

The ML model's reduced freezing rain probability of detection reflects two physically distinct miss regimes operating in opposite thermodynamic directions. Ice pellet misses (24\% of freezing rain observations) occur in environments with stronger freeze layers and higher lower 0°C crossing heights — thermodynamically consistent with ice pellet formation — while rain misses (16\%) occur in environments with weaker freeze layers and warmer surface temperatures physically more consistent with rain (Figure \ref{fig:uncertain}). These opposing thermodynamic signatures mean that no single probability threshold adjustment can simultaneously recover detection in both regimes: lowering the threshold to recover rain misses inevitably reintroduces false alarms in rain-like environments the model currently suppresses correctly, as demonstrated by the threshold analysis in Figure \ref{fig:perf}.
The improved freezing rain success ratio reflects the complementary story. Cases where the ML model correctly predicts rain in environments where area-based methods predict freezing rain verify as rain against mPING 71.2\% of the time and is stable across single-reporter (71.4\%) and multi-reporter (70.6\%) cases. Diagnostic analysis shows a higher level of separation in the environments from confirmed freezing rain environments in surface wet-bulb temperature and freeze layer intensity space but substantially overlap in Bourgouin energy space (Figure \ref{fig:uncertain}, bottom row). This progressive loss of discriminability suggests the ML model exploits surface temperature information available in the raw vertical profile that is not explicitly represented in the area-based framework.

Despite the reduced categorical detection, the ML model's probabilistic output preserves meaningful freezing rain signal in the majority of apparent misses. When the model predicts ice pellets but freezing rain is observed, freezing rain remains the second highest predicted probability in 77\% of cases with a median value of 0.28, compared to a median ice pellet probability of 0.55, indicating genuine uncertainty between the two types rather than confident misclassification. Forecasters examining the full probability distribution rather than the dominant class will therefore have access to freezing rain risk information in most apparent ice pellet miss cases. The approximately 50\% overlap between freezing rain and ice pellet distributions at the 50th percentile density contour across multiple thermodynamic diagnostic spaces confirms that this ambiguity reflects a fundamental atmospheric predictability limit rather than a specific model deficiency.

\subsection{Observational Uncertainty and mPING Reliability}

The reliability of crowd-sourced observations as a verification dataset is an important consideration when interpreting model performance metrics, particularly for freezing rain and ice pellets where the thermodynamic environments are similar and observer discrimination is challenging. The mPING dataset, while offering denser spatial coverage than instrumented networks and uniquely providing ice pellet observations unavailable from automated ASOS sensors, exhibits known limitations including population-based spatial biases, reduced overnight reporting, and variability in observer expertise \citep{MPING}.
Physical quality control based on wet-bulb temperature constraints (Table \ref{tab:qc}) substantially reduced the number of thermodynamically implausible reports and improved both model performance and physical consistency. However, the QC procedure cannot fully resolve observational uncertainty near the thermodynamic boundaries between precipitation types, particularly at the rain/freezing rain interface where surface wet-bulb temperatures are near 0°C and small analysis errors can alter the apparent plausibility of a report. The diagnostic analysis provides direct evidence of this residual uncertainty: cases where the ML model predicts freezing rain but mPING reports rain (Figure \ref{fig:uncertain}, bottom row, blue contours) are thermodynamically indistinguishable from confirmed freezing rain observations in Bourgouin melt/freeze energy space, and are systematically colder at the surface than cases where both the model and observations agree on rain. This spatial alignment suggests that a portion of these apparent false alarms may reflect genuine freezing rain events misreported as rain by mPING observers, rather than systematic model error.
We do not claim that our QC procedure is optimal, and future refinements could improve performance in specific regimes. For example, relaxing the surface wet-bulb threshold for freezing rain could improve probability of detection at the cost of introducing additional observationally uncertain cases into the training and validation datasets. The development of higher-resolution observational networks in areas where mixed precipitation types frequently occur, such as the New York State Mesonet \citep{nym}, could provide more reliable ground truth data for future model development and verification in these challenging regimes.

\subsection{Physical Interpretability}

We evaluate the physical consistency of model behavior through three complementary components that span the full modeling pipeline. First, physically-based quality control of training data ensures the model learns from thermodynamically plausible observations, directly improving both performance and the physical consistency of predictions (Section 3.1). Second, regime-based diagnostic analysis evaluates whether model predictions and errors are physically structured. The thermodynamic analysis in Figure \ref{fig:uncertain} demonstrates that ML prediction errors are not randomly distributed in thermodynamic space but concentrate in physically interpretable regimes consistent with the known difficulty of FZRA/IP discrimination. Third, the interactive visualization tool enables dynamic querying of model predictions through direct profile manipulation, allowing forecasters and model developers to probe model behavior in specific thermodynamic scenarios and identify failure modes that aggregate statistics cannot reveal.

Together these components provide a more complete picture of model trustworthiness than any single approach alone. The quality control procedure establishes that the training signal is physically meaningful; the regime-based diagnostic analysis confirms that the learned relationships are physically structured; and the interactive tool enables ongoing evaluation by domain experts who can test whether model behavior aligns with their meteorological understanding. This approach to physical interpretability is particularly valuable for operational adoption; forecasters are more likely to trust and appropriately use a probabilistic model when they can interrogate its behavior directly rather than relying solely on aggregate verification statistics \citep{Novak2023-th, rogers2023nws}. We note that while we implemented the interactive tool in JavaScript for operational deployment, similar functionality is achievable through simpler frameworks such as Jupyter notebook widgets, making this approach accessible to individual researchers without significant software development overhead.

\subsection{Uncertainty Quantification}

A major drawback of deterministic NWP p-type categorization is the absence of uncertainty quantification from a single model run. Most probabilistic post-processing methods quantify only aleatoric uncertainty, which is expressible directly through a calibrated probability distribution, while epistemic uncertainty typically requires computationally expensive sampling or ensemble strategies (see Appendix A). The evidential neural network framework addresses both simultaneously with computational costs comparable to a standard neural network.
It is important to note that epistemic uncertainty in a data-driven model is not directly comparable to structural uncertainty in a physics-based model. Structural uncertainty in NWP arises from simplifying assumptions, incomplete process representation, and parameterization of subgrid-scale phenomena. In contrast, epistemic uncertainty in an evidential neural network reflects the density of similar training examples in input space: high epistemic uncertainty indicates that the model has encountered few profiles similar to the input and is therefore extrapolating beyond its training distribution. This type of uncertainty is in principle reducible through targeted data collection in underrepresented regimes, and can serve as a practical guide for scientists and model developers identifying where additional observations would most improve model performance. Consistent with this interpretation, the composite profile analysis (Figure \ref{fig:icep_frzr}) showed that high epistemic uncertainty cases were associated with nearly identical FZRA and IP soundings with thermodynamically ambiguous profiles that are likely underrepresented relative to more canonical cases in the training dataset. However, that epistemic uncertainty did not provide additional discriminating power at the individual prediction level beyond what the probability distribution itself conveys in the FZRA/IP ambiguous regime, suggesting its primary value is in identifying broad thermodynamic regimes of limited predictability rather than individual case guidance.

Furthermore, our evidential approach produced well-calibrated probability estimates across all four precipitation types, demonstrating a substantial improvement over the modified area method (Figure \ref{fig:reliability}). When comparing these approaches, it is critical to acknowledge the fundamental difference in their probability spaces. The modified area method operates using independent, marginal probabilities, where the phase equations are evaluated separately, naturally accommodating the occurrence of mixed precipitation. In contrast, the evidential neural network outputs normalized, joint probabilities, treating the primary precipitation types as mutually exclusive classes where the sum of probabilities is constrained to one.

Despite enforcing this mutual exclusivity, the marginal reliability analysis confirms the ML model directly produces highly calibrated probabilities without requiring any post-hoc adjustments. Given this robust calibration and its computational efficiency, the evidential framework is exceptionally well-suited for operational post-processing of NWP output, either as a standalone tool applied to archived model data or embedded directly within an NWP emulation framework. Future work could build upon this by transitioning the neural network to a multi-label architecture, allowing it to explicitly output independent marginal probabilities and natively predict mixed-phase events in a manner more physically consistent with diagnostic methods.
      
\section{Conclusion}

We trained an evidential neural network to predict probabilistic winter precipitation type from NWP vertical profiles, with a focus on physical consistency, uncertainty quantification, and operational interpretability. Rigorous quality control of crowd-sourced mPING observations to remove thermodynamically implausible reports substantially improved model performance and enabled physically consistent predictions across bulk statistical evaluation, case study analysis, and interactive model interrogation.
The ML model outperforms area-based deterministic methods in success ratio for freezing rain and ice pellets while maintaining comparable or better performance for rain and snow. The reduced freezing rain probability of detection reflects two physically distinct miss regimes — ice pellet confusion in strong freeze layer environments and rain confusion in weak freeze layer environments — rather than a uniform model deficiency. In the majority of apparent ice pellet miss cases, freezing rain remains the second highest predicted probability, providing operationally meaningful guidance beyond the dominant class prediction. The model's calibrated probabilistic output appropriately expresses uncertainty in thermodynamically ambiguous regimes where the fundamental discrimination between freezing rain and ice pellets represents a limit of atmospheric predictability rather than a model limitation.
The evidential framework provides epistemic and aleatoric uncertainty estimates from a single model run at computational costs comparable to a standard neural network. Physical interpretability was demonstrated through regime-based diagnostic analysis and an interactive visualization tool that enables direct interrogation of model predictions, supporting both operational forecaster trust and ongoing model development.
Future work includes evaluation beyond the conditional precipitation framework, time-lagged and multi-model ensemble approaches for more robust uncertainty quantification, refinement of the QC procedure, and systematic forecaster feedback through the interactive interface to guide further model development.




\acknowledgments
This material is based upon work supported by the NSF National Center for Atmospheric Research, which is a major facility sponsored by the U.S. National Science Foundation under Cooperative Agreement No. 1852977. This research has also been supported by NSF Grant No. RISE-2019758. We would like to acknowledge computing support from the Casper system (\url{https://ncar.pub/casper}) provided by the NSF National Center for Atmospheric Research (NCAR), sponsored by the National Science Foundation. Interns EK, DK, SR, BS, and JW were hosted at NSF NCAR through the Summer Internships in Parallel Computational Science (SIParCS) program. JR was supported by NOAA Award NA19OAR4320073. We also kindly acknowledge the expert input we received from Phil Schumacher, Mike Fowle, and Andy Just from the National Weather Service; their ideas and expertise helped shaped our model development ideas. 
%
%
\datastatement
Training, validation, and diagnostic data, as well as the model and scaling values, used in this study are available at https://zenodo.org/records/17676792. The MILES-GUESS package is archived at https://doi.org/10.5281/zenodo.10729801, the \textit{ptype-physical} package can be found at https://zenodo.org/records/17677249, and the code for the interactive viewer is located at https://zenodo.org/records/21047319.

%
\appendix
\appendixtitle{Appendix}

\begin{table}
\centering

\caption{As ASOS is capable of outputting mixed types and types that our model doesn't predict, we used the following hierarchy for simplicity, where any label further down the table takes precedent over those above it (represented in the far right colorbar of figures 9 and 12).}
\begin{tabular}{rl}
\hline
ASOS&Label\\
\hline
M&No Precip\\
BR&Mist\\
RA&Rain\\
SN&Snow\\
UP&Unidentified\\
FZRA&Freezing Rain\\
FZFG&Freezing Rain\\
\hline
\end{tabular}
\end{table}





%
\subsection{Epistemic Uncertainty}

Epistemic uncertainty calculated by an evidential neural classifier is quantified by framing classification as a subjective logic problem where the network outputs a non-negative evidence vector $e_k \ge 0$ for each class $k$. For a $K=4$ class problem, this evidence parameterizes a Dirichlet distribution by defining the concentration parameters as $\alpha_k = e_k + 1$. The total Dirichlet strength, $S$, is the sum of these parameters:$$S = \sum_{k=1}^4 \alpha_k$$Epistemic uncertainty, $u$, is then calculated as the total number of classes divided by this total strength:$$u = \frac{K}{S} = \frac{4}{\sum_{k=1}^4 (e_k + 1)}$$In this formulation, a complete lack of evidence ($e_k = 0$ for all $k$) yields a base strength of $S=4$, resulting in maximum epistemic uncertainty ($u = 1$), while accumulating infinite evidence drives the uncertainty toward zero.

\subsection{Aleatoric Uncertainty}

Unlike epistemic uncertainty, aleatoric uncertainty can be estimated directly from probabilities. The formula for total aleatoric uncertainty is $\sum p_{\mathrm{all}} (1 - p_{\mathrm{all}})$, where $p_{all}$ is the array of probabilities for all classes for a given sample.

%
%
\nocite{*}
\bibliographystyle{ametsocV6}
\bibliography{ptype}

%

%

\end{document}